\documentclass[preprint2]{aastex}
\usepackage{lscape}
\usepackage{graphicx}
\usepackage{epsf}

\shorttitle{The B and Be Stars of $h$ and $\chi$ Per}
\shortauthors{Boyer, McSwain, Aragona, \& Ou-Yang}
\slugcomment{Submitted to AJ on 16/6/2012}

\begin{document}

\title{Physical Properties of the B and Be Star \\
	Populations of $h$ and $\chi$ Persei}

\author{Amber N. Marsh Boyer\altaffilmark{1}, M. Virginia McSwain\altaffilmark{1,2}, Christina Aragona\altaffilmark{1}, \\and Benjamin Ou-Yang\altaffilmark{3}}
\affil{Department of Physics, Lehigh University, 16 Memorial Drive East,
    Bethlehem, PA 18015}
\email{anm506@lehigh.edu, mcswain@lehigh.edu, cha206@lehigh.edu, ouyang@chara.gsu.edu}

\altaffiltext{1}{Visiting Astronomer, Kitt Peak National Observatory, National Optical Astronomy Observatory, which is operated by the Association of Universities for Research in Astronomy (AURA) under cooperative agreement with the National Science Foundation.}
\altaffiltext{2}{The WIYN Observatory is a joint facility of the University of Wisconsin-Madison, Indiana University, Yale University, and the National Optical Astronomy Observatory.}
\altaffiltext{3}{Current address: Department of Physics and Astronomy, Georgia State University, P.O. Box 4106, Atlanta, GA 30302}

\begin{abstract}
	We present a study of the B and Be star populations of the Double Cluster $h$ and $\chi$ Persei. Blue optical spectroscopy is used to first measure projected rotational velocity, $V$ sin $i$, effective surface temperature, $T_{\rm eff}$, and surface gravity, log $g$, for B-type sample stars, while available Str\"{o}mgren photometry is used to calculate $T_{\rm eff}$ and log $g$ for the Be stars showing emission. In our sample of 104 objects for which we measured these stellar parameters, 28 are known or proposed Be stars. Of these Be stars, 22 show evidence of emission at the times of our observations, and furthermore, we find evidence in our data and the literature for at least 8 transient Be stars in the clusters. We find that the Be stars are not rotating near their critical velocity, contrary to the results of studies of similar open clusters. We compare the results of our analysis with other previous studies, and find that the cluster members are more evolved than found by Huang \& Gies but still retain much of their initial rotational angular momentum.
\end{abstract}


\section{Introduction}
NGC 869 and NGC 884 ($h$ and $\chi$ Persei, respectively) are a well known double open cluster, visible in the Northern hemisphere, and have been the focus of many studies over the years. The early 1900's saw a number of studies attempting to determine cluster membership, positions, and radial velocities \citep{Messow1913, Adams1913, Hertzsprung22}. By the 1960's more extensive studies, such as that of \cite{Slettebak68}, were being conducted to determine spectral types for the cluster constituents. More recently an extensive study has been conducted by \citet{Currie10} in which they investigated the general properties and membership of the clusters. Their results, in agreement with those of \citet{BK05} and \citet{Slesnick02}, find that the clusters are incredibly similar, having common ages of $\sim$13--14 Myrs, distance moduli dM $=11.8-11.85$ ($\sim 2,200$ pc), and reddenings of E($B-V$) $\sim 0.52-0.55$. They also estimate a total mass of at least 20,000 M$_\odot$ for the clusters.

One of the prominent motivations for our study is that these young open clusters are rich in Be stars. As early as the 1920's, observational studies conducted by \citet{Trumpler26} and others noted the presence of emission in the hydrogen lines of many of the brightest B-type cluster members. Modern studies of the cluster have shown that upwards of 30\% of the brightest B-type stars are known to be Be stars \citep{Keller01}. In a study conducted with \emph{Spitzer}, \citet{Currie08} investigated the lower mass stellar population for mid-infrared excesses due to the presence of protoplanetary disks. They also identified 57 Be stars and candidates exhibiting excess emission at 24$\mu$m, which helped to motivate our study to follow up these candidates and confirm their Be nature. Of their stars, 21 had previously been identified as showing emission, and 20 of their stars are included in the present study. 

The modern working definition of a Be star is given as ``a non-supergiant B star whose spectrum has or had at some time, one or more Balmer lines in emission" \citep{Porter03}. While this general definition also encompasses objects such as the well known Algol binary systems and Herbig Ae/Be stars, classical Be stars are further delineated as having circumstellar line emission formed in an optically thin equatorial disk, low-order line profile variations, and rapid rotation \citep{Porter03}. These Be disks are comprised of warm gaseous material ejected from the stellar surface during outburst events. The gas is then pulled into a gravitationally bound orbit about the stellar equator.

It is well established that as a population, Be stars rotate faster than than their non-emission, B-type counterparts. Precisely why this is the case, however, is still debated. There are three primary theories as to why Be stars are rapid rotators: they may have been born as rapid rotators, spun up by mass transfer in a close binary system, or spun up during the main sequence evolution of B-type stars. The observed rotation rates of Be stars are $\ge$60--80\% of their critical velocity \citep{McSwain08}, at which point the gravitational and centrifugal forces are balanced, although recent results suggest that this threshold may be mass dependent \citep{Huang10}. The main sequence lifetimes of these objects are likely extended as a direct result of their rapid rotation, as this fosters rotational mixing of their stellar interiors and replenishes their hydrogen cores \citep{Meynet00}. However, rapid rotation alone is not enough to spur the photospheric material of these stars to form the disk structures they host. It is likely that other weaker processes, such  as non-radial pulsations (NRPs), are needed to provide the additional angular momentum necessary for this material to leave the stellar surface \citep{Rivinius01,Porter03,McSwain08,Cranmer09}. A growing number of Be stars have been identified to exhibit NRPs (see \citealt{Rivinius03,Emilio10}). 

B-type and Be stars have become important targets for asteroseismology. In hot stars, different pulsation frequencies provide information about different layers of the stellar interior. The information gleaned from studies targeting B stars with NRPs is being used to improve current stellar structure and stellar evolution models. Doing so, however, requires accurately determined stellar surface parameters in order to set appropriate boundary conditions for these models. Accurate measures of effective temperature and surface gravity are particularly essential to determining stellar radii for these models, as well as for determining stellar ages and evolutionary spin-down. Many B-type and Be stars in $h$ and $\chi$ Per have been found to host NRPs, and currently there are on-going campaigns to observe and characterize the nature of the variable pulsations found in these stars \citep{Krzesinski97, Krzesinski99, Saesen10}. The interest in the pulsating stars of $h$ and $\chi$ Per necessitates improved measurements of stellar parameters for the cluster.

There is an on-going debate in the massive star community regarding the evolution of angular momentum of B-type stars. With their abundance of B-type stars, $h$ and $\chi$ Per are two of the many stellar clusters at the center of this debate. \citet{Strom05} find that the present-day rotation rates of these stars are set by environmental characteristics of the natal clouds in which they formed, with little change over the main sequence stellar lifetime. The work of \citet{HG06} and \citet{Huang10}, however, indicates that the observed rotation rates of B-type stars are due less to the initial birth-line rotational rates of the stars and more to evolutionary spin-down or mass transfer in binaries. Both studies also observe that B-type stars in clusters have, on average, significantly higher rotation rates than field B-type stars. Resolving the angular momentum problem also requires a new examination of the stellar parameters and evolutionary states.

In this work we present stellar properties determined from blue optical spectroscopy of a sample of B and Be stars in the clusters. Section 2 provides details of spectroscopic observations taken with various instruments. In Section 3 we discuss our methods for determining $V$ sin $i$, $T_{\rm eff}$, and log $g$ for both the B and Be star populations.  We feel that in order to study Be stars relative to their non-emission peers it is important to apply a uniform technique for B and Be star analysis (in as much as possible). With this homogenous and independently determined dataset we will be able to further investigate the nature of the clusters' Be star population. Section 4 quantitatively compares our results with those of \citet{HG06} and \citet{Strom05}, who use different methods to determine the same physical parameters. We discuss the impact of our disagreements in the context of the evolution of stellar angular momentum. Finally, in Section 5 we draw our conclusions regarding the clusters' massive star populations. We plan to use the results of this work to examine the spectral energy distributions (SEDs) of the star + disk systems among the Be star population in a forthcoming paper.

\section{Observations}

We have obtained spectra for a total of 104 members of NGC 869 and NGC 884 during multiple observing runs:  2005 November using the Kitt Peak National Observatory (KPNO) Wisconsin Indiana Yale NOAO (WIYN) 3.5 m telescope with the Hydra multifiber spectrograph;  2010 August using the Wyoming Infrared Observatory (WIRO) 2.3 m telescope with the Long Slit spectrograph; 2011 November using the KPNO 2.1 m telescope with the GoldCam spectrograph; and 2012 January using the 0.9 m KPNO Coud\'e Feed (CF) telescope with the Coud\'e spectrograph. The UT dates, wavelength range, resolving power, number of targets, and instrumental setup details for all runs are summarized in Table \ref{spectroscopy}.

All of the spectra obtained at the WIYN 3.5 m with the Hydra spectrograph have been zero corrected using standard routines in IRAF\footnote[4]{ IRAF is distributed by the National Optical Astronomy Observatory, which is operated by AURA, Inc., under cooperative agreement with the NSF.}, and have been flat-fielded, wavelength-calibrated, and sky-subtracted in IRAF using the \textbf{dohydra} routine. 
The Hydra observations obtained by M. Virginia McSwain in 2005 consist of 7 exposures of NGC 869, totaling 2.25 hrs, and 5 exposures of NGC 884, totaling 2 hrs. Each exposure has been wavelength calibrated via a CuAr comparison spectrum both before and after the cluster observations. For each of the two configurations, the exposures have been transformed to a common heliocentric wavelength grid and co-added to produce good signal-to-noise for each star. The spectra were rectified to a unit continuum by fitting line-free regions. 

Before any reduction or calibration routines were applied to the WIRO 2.3 m data, all spectra (object and comparison) were corrected for bit-flip errors with the \textbf{rfits} routine in IRAF. A CuAr calibration lamp source was used to obtain wavelength calibration spectra before and after every object spectrum. The spectra were then zero-corrected, flat-fielded, wavelength-calibrated, and rectified to a unit continuum using standard slit spectra routines in IRAF. Fainter objects that required multiple exposures were co-added prior to continuum rectification to improve signal-to-noise. Given the poorer resolution of blue WIRO spectra, we cannot use these data with our spectral fitting techniques described in Section 3. We therefore only use the blue and red WIRO spectra to identify Be stars exhibiting emission at the time of our observations.

Spectra from the KPNO 2.1 m instrument have been similarly zero-corrected, flat-fielded, and wavelength-calibrated using the standard routines found in IRAF. Once wavelength calibrated via the HeNeAr comparison lamp spectra, which were taken before and after every object spectrum, the data were rectified to a unit continuum. 

In a similar manner, the spectra obtained with the 0.9 m KPNO CF telescope have been zero-corrected, flat-fielded, and wavelength-calibrated using the standard routines in IRAF. ThAr comparison spectra were taken every one to two hours during the run. The data were then rectified to a unit continuum.

\section{Physical Parameters from Spectral Models}
The first component of our population study is to determine basic parameters for the cluster constituents using a methodology devised by \citet{McSwain08}. Using ground based optical spectroscopy (4000--5200 \AA) and model fitting techniques, we can determine $V$ sin $i$, $T_{\rm eff}$, and log $g$ for each star. To obtain these measurements, we compare our observed spectra to grids of model B-type stars, determining a best fit to the data by minimizing the mean square of the deviations rms$^2$. 

\begin{figure}[h]
\hspace{-.75in}
\includegraphics[angle=90,width=.6\textwidth]{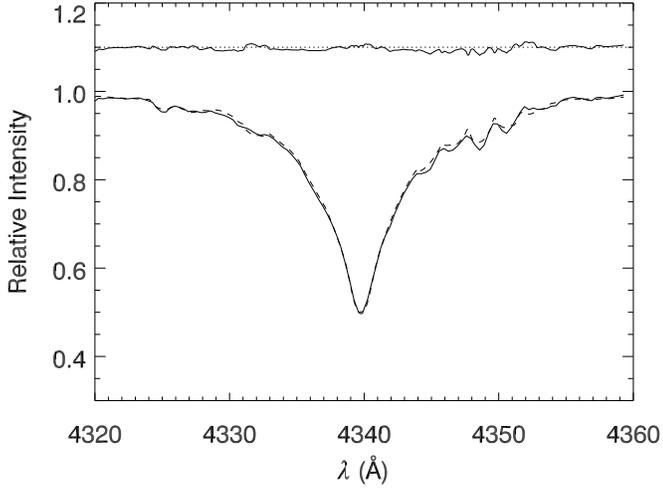}
 \vspace{.1in}
 \hspace{-.75in}
 \includegraphics[angle=90,width=.6\textwidth]{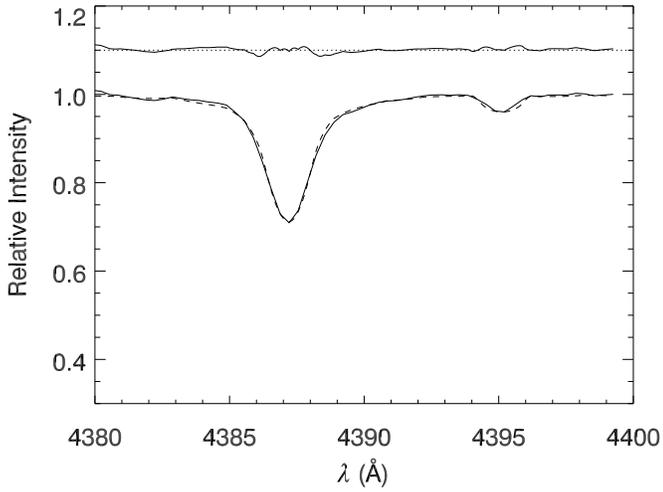} 
\caption{Sample spectral line fits for NGC 869-90. Shown on the top is H$\gamma$ and on the bottom is He I  $\lambda$4378. The solid line is our observed spectrum while the dashed line displays our model fit to the line, with the computed residual shown above, shifted for clarity. }
\label{specfits}
\end{figure}

We begin our analysis by making a rough estimate of $T_{\rm eff}$ and log $g$ for a star, and then we compare the He I $\lambda\lambda$4387, 4471, 4713, and Mg II $\lambda$4481 lines  with the Kurucz ATLAS9 models \citep{Kurucz94} to determine $V$ sin $i$. These lines are used because their broadening is dominated by the effects of rotation, thus yielding a better indication of $V$ sin $i$. We then take a weighted average of the four values determined from each of the lines to give our measured value of $V$ sin $i$. The error, $\Delta V$ sin $i$, is determined by the offset from the measured best-fit value that increases the rms$^2$ by 2.7rms$^2/N$. Here, $N$ is the number of wavelength points within the fit region. A sample fit  determined for He I $\lambda$4387 in NGC 869--90 is shown in the bottom panel of Figure \ref{specfits}. Our results for $V$ sin $i$ and its errors are listed in columns 2 and 3 of Tables \ref{starparams} and \ref{Be_params}.

In Be stars it is possible that the He I lines may contain weak emission from the circumstellar disk, which may partially fill the absorption features and narrow the overall line profile. So while we have made measurements of  $V$ sin $i$ where we can for Be stars in our sample, these values should be considered as lower limits. 

We measure a mean $V$ sin $i=$157 km s$^{-1}$ with a standard deviation of 89 km s$^{-1}$ for the normal B-type stars of both clusters, including binary systems. Assuming an average inclination angle of $i=60^o$, this gives a mean $V_{\rm eq}=181$ km s$^{-1}$ for our sample of B stars. For the Be stars we measure a mean $V$ sin $i=205\pm 81$ km s$^{-1}$, with a mean $V_{\rm eq}=237$ km s$^{-1}$. From this it is clear that the Be stars in these clusters are, on average, rotating somewhat more rapidly than their B-type counterparts. 

\begin{figure}[h]
\hspace{-.75in}
 \includegraphics[angle=90,width=.6\textwidth]{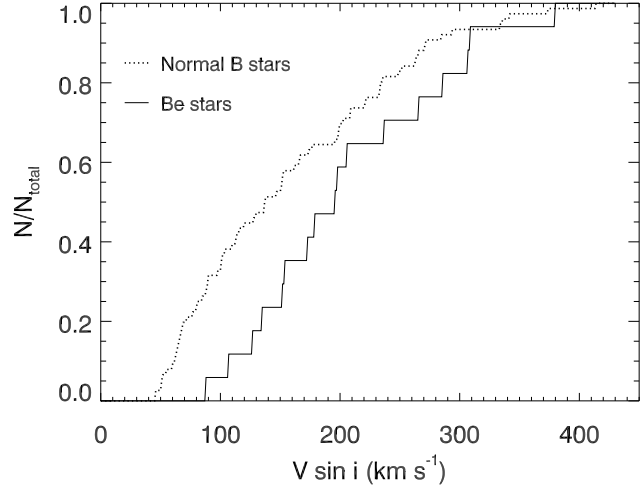} 
\caption{ Cumulative distribution function of $V$ sin $i$ for the Be stars (\emph{dashed line}) and the normal B-type stars (\emph{dotted line}) of both NGC 869 and NGC 884.}
\label{Vdistrib}
\end{figure}

The cumulative $V$ sin $i$ distributions for all Be and normal B-type stars in both NGC 869 and NGC 884 are shown in Figure \ref{Vdistrib}. In comparison to other young open clusters (see \citealt{McSwain08,McSwain09}), we find that the Be star population of NGC 869 and NGC 884 are rotating surprisingly more slowly that expected in comparison to their B-type counterparts. Using the two-sided Kolmogorov-Smirnov (K-S) statistical test, we investigate the null hypothesis that the distributions of B-type and Be stars differ. The K-S test indicates a 7.6\% chance that the two populations are drawn from the same sample. Using the values of mass, $M_{\star}$, and radius, $R_{\star}$, discussed below for all stars in our sample, we can determine the critical velocity \begin{eqnarray}
	V_{\rm crit} = \sqrt{\frac{G M_{\star}}{R_{\rm e}}} \label{Vcrit}
\end{eqnarray} for our stars. For simplicity in this expression, we assume that the polar radius of the star, $R_{\rm p}$, is equal to $R_{\star}$, and that a rotationally distorted star has an equatorial radius $R_{\rm e} = 1.5 R_{\rm p}$. With this, we find a mean $V_{\rm{crit}}$ of 430 km s$^{-1}$ for the B-type and Be stars in these clusters.

Having determined values of $V$ sin $i$ for each star, we turn again to model spectral fitting to determine values for $T_{\rm eff}$ and log $g$ from the H$\gamma$ line at 4340\AA. The hydrogen Balmer lines are particularly sensitive to $T_{\rm eff}$ and log $g$, making them ideal for determining these quantities accurately. The method outlined here was employed for normal B-type stars and Be stars with no emission present in this work. For stars having $T_{\rm eff}$ $\le$ 15000 K we employ the methods of \citet{HG06}, who use H$\gamma$ line profiles generated by the line-blanketed, local thermodynamic equilibrium (LTE) Kurucz ATLAS9 and SYNSPEC codes. The ``virtual star" models produced by their code simulate spherically symmetric stars with constant  $T_{\rm eff}$ and log $g$ across their surface, and the spectra of these model stars are then used to measure these values and their errors of our observed spectrum, similar to our procedure for $V$ sin $i$. The errors $\Delta T_{\rm eff}$ and $\Delta$log $g$ are determined from the quadratic sum of the $V$ sin $i$ propagated errors and the errors due to the intrinsic noise in the observed spectrum. For hotter stars, LTE models should systematically underestimate $T_{\rm eff}$ as non-LTE effects alter the equivalent width of the H$\gamma$ line we are measuring. Hence for stars having $T_{\rm eff}$ $\ge$ 15000 K, we use instead the metal line-blanketed, non-LTE, plane-parallel, hydrostatic TLUSTY BSTAR2006 model spectra \citep{LH07}. A sample fit of the H$\gamma$ line in NGC 869--90 is shown in the top panel of Figure \ref{specfits}. The errors $\Delta T_{\rm eff}$ and $\Delta$log $g$ are determined from the values that produce an rms$^2$ no more than 2.7rms$^2/N$ greater than the minimum rms$^2$. Our results for $T_{\rm eff}$, log $g$, and their respective errors are shown in columns 4--7 of Table \ref{starparams}.

29\% of the stars in our sample are rapid rotators, having measured values of $V$ sin $i$ in excess of 200 km s$^{-1}$. At such significant rotational velocities, assuming a spherical shape for these stars is no longer plausible given the substantial centrifugal forces distorting the stars into oblate spheroids. This rotational distortion produces significant differences in both the temperatures and surface gravities at the polar and equatorial regions. As the measured values of  $T_{\rm eff}$ and log $g$ are averages across the visible stellar hemisphere, these rotational effects produce lower values than expected, a phenomenon known as gravitational darkening. As the rotation rate for a star approaches its critical velocity, the equatorial radius may increase by as much as 50\%, while the polar radius remains unchanged. For these reasons, we convert our measured log $g$ to log $g_{\rm polar}$, as detailed in \citet{HG06}. The authors produce detailed spectroscopic models to investigate the effects of such rotational distortions and determine a statistical correction factor for log $g$. This factor is averaged over all possible values of inclination angle, $i$, for a variety of stellar models. By bilinearly interpolating between their models, we converted our measured value of log $g$ to log $g_{\rm polar}$. This value of  log $g_{\rm polar}$ is a better measure of the true surface gravity of the star, untainted by the effects of rapid rotation, and provides us a more accurate means of comparison between slowly and rapidly rotating stars. The log $g_{\rm polar}$ correction is therefore most significant for stars rotating more rapidly than 50\% of $V_{\rm{crit}}$. We assume that this conversion produces a negligible change to the value determined previously for $\Delta$log $g$. Our derived values of log $g_{\rm polar}$ are listed in column 8 of Table \ref{starparams}. 

\begin{figure}[h]
\hspace{-.75in}
 \includegraphics[angle=90,width=0.6\textwidth]{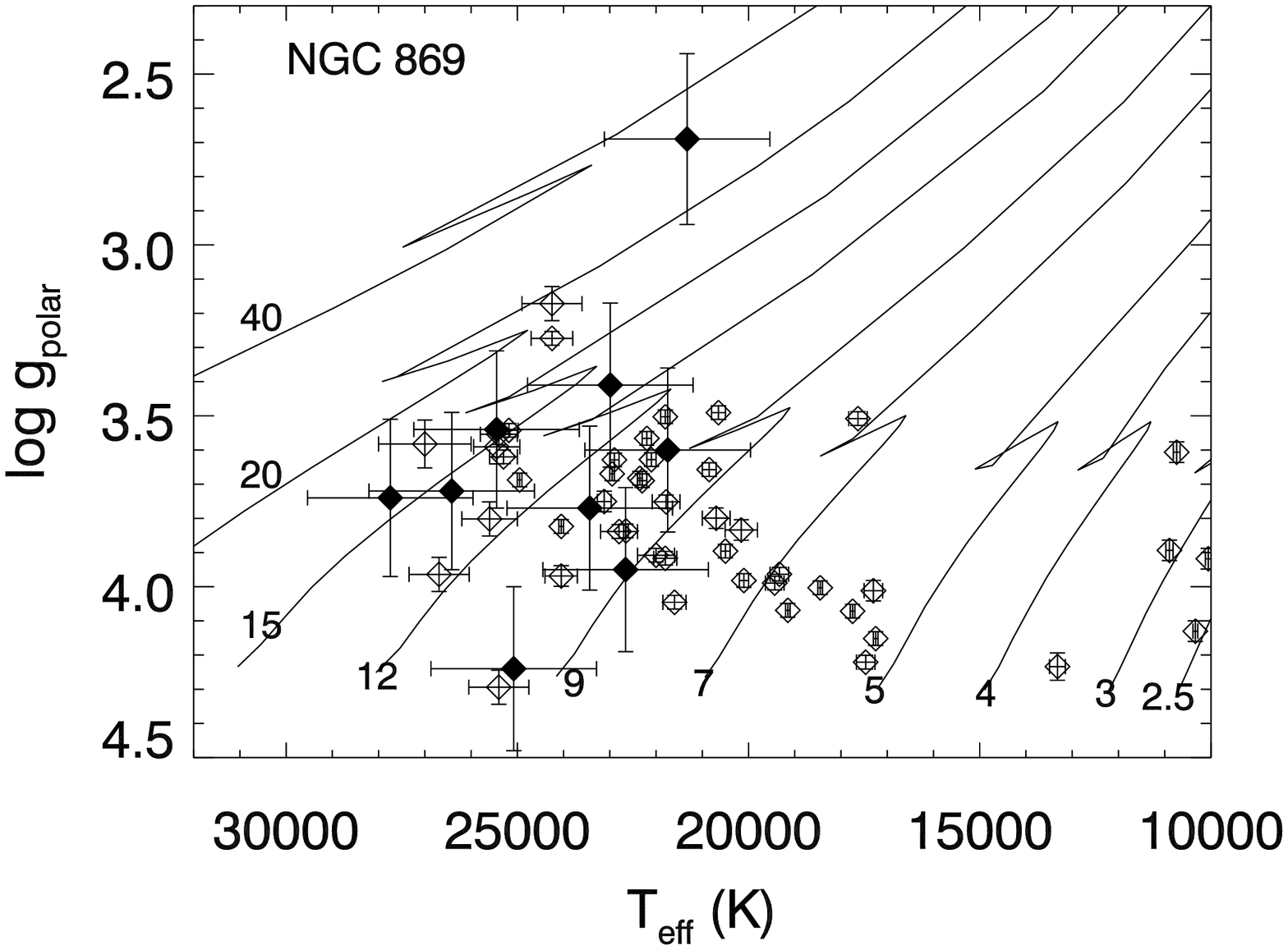}
 \vspace{.1in}
 \hspace{-.75in}
  \includegraphics[angle=90,width=0.6\textwidth]{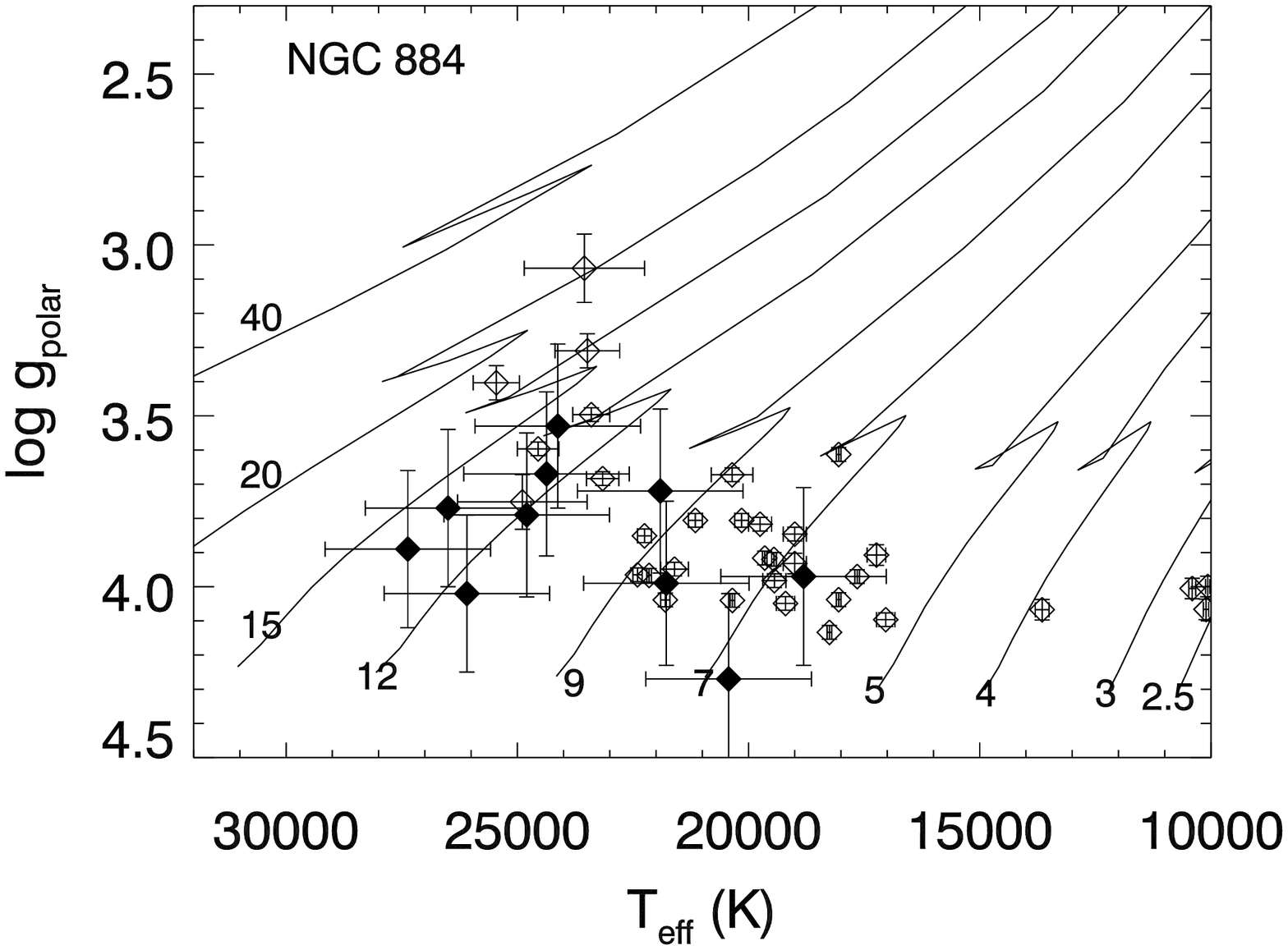} 
\caption{ For both NGC 869 (top) and NGC 884  (bottom), $T_{\rm eff}$ and log $g_{\rm polar}$ are plotted with the evolutionary tracks of \citet{Schaller92}. The zero age main sequence (ZAMS) mass of each evolutionary track is labeled along the bottom. Normal B-type stars are shown as open diamonds while Be stars are filled diamonds. }
\label{evolution}
\end{figure}

Once we have measured parameters for all of the B-type stars, we can then determine $M_{\star}$ and $R_{\star}$ for each of them by interpolating values from the \citet{Schaller92} non-rotating evolutionary tracks, consistent with the slow rotation of most of our targets. These evolutionary tracks are shown plotted with $T_{\rm eff}$ and log $g_{\rm polar}$ in Figure \ref{evolution}.  The errors $\Delta M_{\star}$ and $\Delta R_{\star}$ correspond to our measured $\Delta T_{\rm eff}$ and $\Delta$log $g$. Additionally, we have compared our results with the rotating models of \citet{Ekstrom12}, and we find agreement between the models to within 10\%. The resulting values of $M_{\star}$, $R_{\star}$, and their respective errors are listed in columns 9--12 of Tables \ref{starparams} and \ref{Be_params}. We have also checked the accuracy of our results by comparing the TLUSTY BSTAR2006 model SEDs  with the observed SEDs for our B-type sample stars, and find excellent agreement between our derivations of distance and reddening with the accepted values of \citet{Currie10}. These results will be discussed further in a forthcoming paper.

We note that star NGC 869--566 did not show any signs of emission in our initial observations, but has since exhibited progressively stronger emission in our more recent 2010 and 2012 observations; hence we include it among the non-emission Be stars and have measured $T_{\rm eff}$ and log $g$ from the H$\gamma$ line it its 2005 blue spectrum. For the other Be stars in our sample we cannot determine accurate values for $T_{\rm eff}$ and log $g$ via the same model fitting technique, as hydrogen absorption line profiles are altered by emission during disk phases of these stars. Instead we can use Str\"{o}mgren photometry  available from the WEBDA\footnote[5]{ Available online at www.univie.ac.at/webda and maintained by Ernst Paunzen} database to correlate $T_{\rm eff}$ and log $g$ for all of the B-type and Be stars in our sample. Non-emission B stars with both available Str\"{o}mgren photometry and spectral model fits were included as calibration stars for the photometric technique. To this we add eight additional stars from \citet{Napiwotzki93} with well-known $T_{\rm eff}$ and available Str\"{o}mgren data. We use \emph{ubvy} magnitudes to first determine the Str\"{o}mgren indices $m_{1}$ and $c_{1}$, given by
\begin{eqnarray}
	m_{1}=& (v-b)-(b-y)\label{m1}\\
	c_{1}=& (u-v)-(v-b)\label{c1}.
\end{eqnarray}
The calculation of $T_{\rm eff}$ determined by \citet{Balona84} requires us to convert the $c_1$ index to the dereddened index $c$ via the expressions  
\begin{eqnarray}
	E(b-y)=& E(B-V)\times 0.754\label{red}\\
	c_{0}=& c_{1} - 0.19[E(b-y)]\label{c0}\\
	c=& \log (c_{0} + 0.200),\label{c}
\end{eqnarray}
which use the reddening values to the clusters, E($B-V$) $=$ 0.55 and E($B-V$) $=$ 0.52 for $h$ and $\chi$ Per, respectively \citep{Currie10,BK05}. Using these indices and the H$\beta$ line magnitude ($\beta$), we can then calculate $T_{\rm eff}$ via the relationship 
\begin{eqnarray}
	\log  T_{\rm Balona} = 3.9036-0.4816(c)-0.5290(\beta) \nonumber \\
	{}-0.1260(c)^2+0.0924(\beta)(c)-0.4013(\beta)^2\label{balonarelation}
\end{eqnarray} 
given by \citet{Balona84}.

\begin{figure}[h]
\hspace{-.75in}
\includegraphics[angle=90,width=0.6 \textwidth]{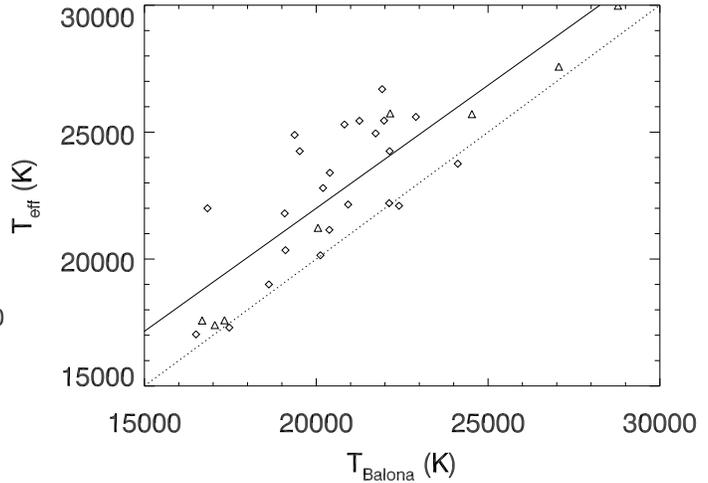} 
\caption{$T_{\rm eff}$ measured for the B-type temperature calibration stars from our work (\emph{diamonds}) and from \citet{Napiwotzki93}(\emph{triangles}) compared to their calculated $T_{\rm Balona}$ \citep{Balona84}. A linear fit of the two temperature scales (\emph{solid line}) and the 1:1 agreement (\emph{dotted line}) are also shown. }
\label{temp cal}
\end{figure}

\citet{McSwain08} found that this calculated value, $T_{\rm Balona}$, slightly underestimates the true $T_{\rm eff}$ of the B-type stars. Thus, we performed a linear fit to the data, shown in Figure \ref{temp cal}, and determined a correction factor that will bring the two independent measurements into agreement. In this way, we are able to use the B-type stars measured by both methods as a calibration to yield values of $T_{\rm eff}$ for the otherwise immeasurable Be stars. These $T_{\rm eff}$ and our calculated errors are listed in columns 4 and 5 of Table \ref{Be_params}. 

\begin{figure}[h]
\hspace{-.75in}
\includegraphics[angle=90,width=0.6 \textwidth]{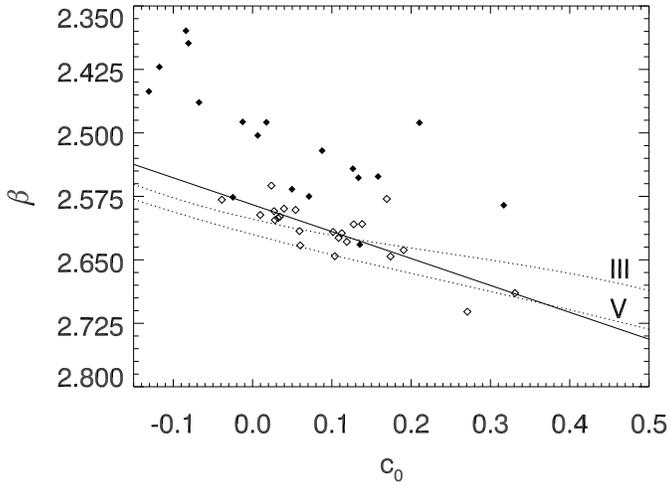} 
\caption{Str\"{o}mgren $c_0$ index and $\beta$ magnitude are plotted with the $c_0 - \beta$ relations for luminosity class V and III stars (\citealt{BS84}; \emph{dotted lines}). The B-type temperature calibration stars from this work (\emph{diamonds}) are plotted with fifteen Be stars (\emph{filled diamonds}) to demonstrate that the Be star $\beta$ magnitudes are brightened due to the disk emission present. A best fit line for the calibration stars is also shown (\emph{solid line}).}
\label{beta cal}
\end{figure}

To determine log $g$, \citet{Balona84} advocate using $\beta$ and $c_0$. Given that $\beta$ serves as an indicator of log $g$ via spectra line width, and $c_0$ is an indicator of temperature in hot stars, the relationship between these two values can serve as a Hertzsprung-Russell diagram for the cluster providing a means for investigating the temperature and evolutionary trends of the stellar population. In more evolved giant or supergiant stars (luminosity classes III and I, respectively), the hydrogen lines are very narrow due to the lower densities in the outer atmospheres of these stars, decreasing the collisional rates that produce the pressure broadening mechanism. Main sequence stars (luminosity class V), which host more dense atmospheres and thus higher collisional rates and higher pressures, have broader hydrogen lines. In general, we do see that the values of $c_0$ and $\beta$ for the B-type calibrators shown in Figure \ref{beta cal} agree with the relations for class V and III stars of \citet{BS84}. However, the circumstellar disks present in Be stars (shown as filled diamonds in Figure \ref{beta cal}) will artificially brighten the $\beta$ magnitudes of these stars, contaminating the $c_0 - \beta$ relation for these stars and our calculated values of log $g$. The spread in log $g$ of the B-type calibration stars prevents us from simply applying either of the $c_0 - \beta$ relations shown in Figure \ref{beta cal} to the population. Instead, we perform a linear fit to the calibration star data and obtain a corrected value of $\beta$ which we then use to correct our calculated $T_{\rm Balona}$ and finalize our calibrated, calculated temperatures and surface gravities for the Be stars exhibiting emission in our data. Since the clusters are approximately the same age, this single fit is appropriate. Additional details regarding the method to determine $T_{\rm eff}$, log $g$, $M_{\star}$, $R_{\star}$, and their respective errors for the Be stars can be found in \citet{McSwain08}. We do not perform any further correction to obtain log $g_{\rm polar}$ for the Be stars measured with this technique, given the large scatter between the calibrators' log $g_{\rm polar}$ and their Str\"{o}mgren log $g$. The final results for log $g$ and $\Delta$log $g$ of the Be stars are listed in columns 6--7 of Table \ref{Be_params}. 

We highlight our results regarding the Be star population in Table \ref{Be_params}, and we note that these Be stars were selected from the literature based on previous observations of spectral line emission or via photometric surveys needing spectroscopic confirmation. Presented within Table \ref{Be_params} are a total of 28 known or proposed Be stars within our sample of 104 stars. For these objects we were able to examine the state of emission from our observations, and then determine their stellar parameters by our spectral modeling or Str\"{o}mgren photometry methods as detailed earlier in this section. The broad-band Str\"{o}mgren indices used to derive their $T_{\rm eff}$ and log $g$ are not likely to be affected by their rapid rotation. We do see evidence of emission in 22 of these Be stars. Interestingly, we see evidence in our data for at least 8 ``transient" Be stars \citep{McSwain08}. The stars NGC 869--146, NGC 869--717, NGC 869--1268, and NGC 884--2262 were observed as Be stars in the past \citep{Slettebak85,Fabregat94,Keller01,BK02}, and the star NGC 884--2468 is a proposed candidate Be star \citep{Currie08}. However, we do not see evidence of emission in our observations of these objects. As mentioned previously, the known Be star NGC 869--566 was initially observed by us in 2005 and showed no sign of emission in its spectrum. However, it has since developed increasingly stronger emission in our 2010 and 2012 observations. For the other two transient Be stars (NGC  869--49, NGC 884--1772), we do see emission in our spectra, however other authors have noted them in non-emission phases in the past \citep{Schild66,Slettebak85, BK02, Keller01}. We are currently preparing a followup paper that discusses the H$\alpha$ emission properties of all the Be stars in our sample. 

\section{Comparison with Other Studies}
Previous studies, such as \citet{Strom05} and \citet{HG06}, have investigated some of our B star targets to determine the same basic stellar parameters we do, however their analyses employed LTE atmospheric models or quantitative corrections to LTE model measurements to account for non-LTE effects. Between these two studies there are clear discrepancies in their measurements and the conclusions each draws regarding the natal rotation rates and angular momentum evolution of the clusters' massive stars. An accurate determination of a given star's surface gravity is essential to the evaluation of stellar radius and evolutionary state. As we show here, non-LTE effects can contribute to significant errors in measurements of log $g$. 

\begin{figure}[h]
\hspace{-.75in}
\includegraphics[angle=90,width=0.6 \textwidth]{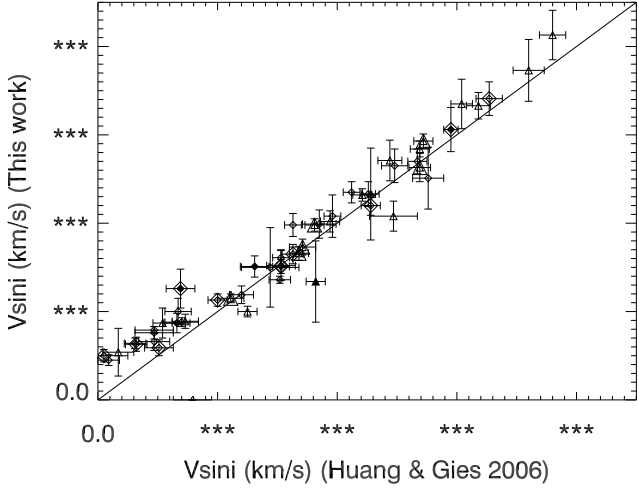} 
\caption{ Comparison of our $V$ sin $i$ measurements with those of \citet{HG06}. NGC 869 cluster members are shown as open diamonds, while NGC 884 members are shown as open triangles. Be stars are shown as filled diamonds and triangles. Spectroscopic binaries, as noted in Tables \ref{starparams} and \ref{Be_params}, are highlighted by double-sized symbols.}
\label{vsinicomp}
\end{figure}

Amongst the 54 stars common to our sample and that of \citet{HG06}, we find some discrepancy between our respective results. As can be seen in Figure \ref{vsinicomp}, there is very good agreement in our determinations of $V$ sin $i$. This is to be expected since we both used LTE models to fit the He I and Mg II lines to measure $V$ sin $i$. The differences at low $V$ sin $i$ may be due to a difference in spectral resolution between our respective datasets. 

\begin{figure}[h]
\hspace{-.75in}
\includegraphics[angle=90,width=0.6 \textwidth]{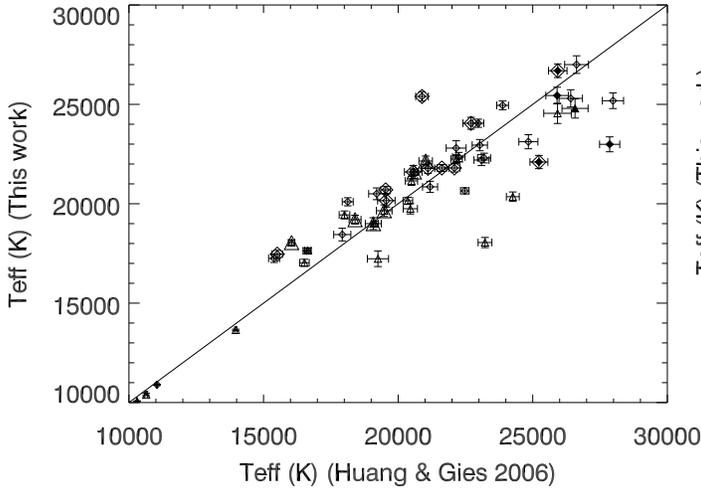} 
\caption{ Comparison of our $T_{\rm eff}$ measurements with those of \citet{HG06}, in same format as Figure \ref{vsinicomp}.}
\label{tempcomp}
\end{figure}

\begin{figure}[h]
\hspace{-.75in}
\includegraphics[angle=90,width=0.6 \textwidth]{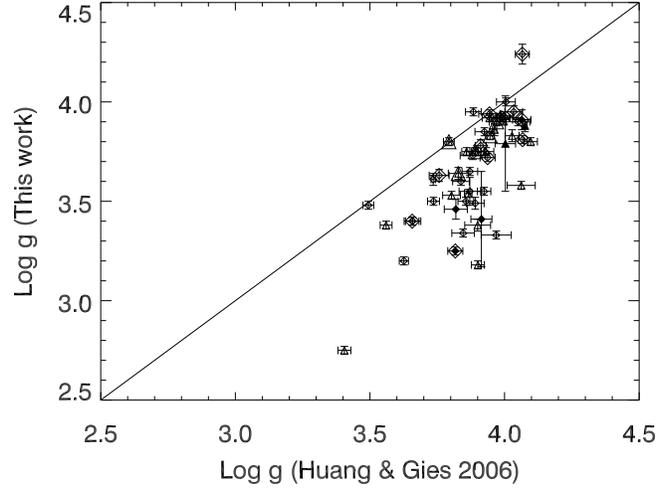} 
\caption{ Comparison of our log $g$ measurements with those of \citet{HG06}, in same format as Figure \ref{vsinicomp}.}
\label{loggcomp}
\end{figure}

However, differences in our measured temperatures for stars with $T_{\rm eff}\ge 15000$ K and in our measured log $g$ values are clearly apparent in Figures \ref{tempcomp} and \ref{loggcomp}, respectively. Given that our general methodology for measuring $T_{\rm eff}$ and log $g$ is the same as \citet{HG06}, the source of these discrepancies lies partially in our use of different stellar models. While \citet{HG06} employ the LTE Kurucz ATLAS9 models \citep{Kurucz94}, we use the more recently available non-LTE TLUSTY BSTAR2006 models of \citet{LH07} for those stars in our sample with $T_{\rm eff}\ge 15000$ K. In their analysis \citet{HG06} acknowledge that their derived temperatures are likely to be slightly lower and gravities slightly higher compared to measurements derived from non-LTE model atmospheres, as shown by the comparative analysis of \citet{LH07}.

\citet{LH07} find that in the non-LTE models the hydrogen Balmer lines tend to be broader and stronger due to the overpopulation of the $n=2$ energy state, thus LTE models will yield overestimated surface gravities due to the altered the shape of the Balmer line wings. \citet{Przybilla11} compared LTE ATLAS9 models to non-LTE TLUSTY models for temperatures between 15000--35000 K.  They found that non-LTE effects are significant above 22,000 K, affecting both the cores and wings of the Balmer lines.  They find that LTE Balmer line profiles have equivalent widths up to 30\% lower than in non-LTE line profiles. This would cause a non-LTE model to find a higher temperature for the same observed line, or LTE models underestimate the temperature. \citet{Przybilla11} also find that LTE models of the H$\gamma$ line may overestimate log g by up to 0.2 dex. 

 While the expected temperature disagreement is opposite of the trend we find in our comparison of our work with \citet{HG06} results, several of our common sample stars with temperatures greater than 24,000K are Be stars, some of which we find to exhibit transient behavior. It is likely that emission has subtly filled in or otherwise altered the H$\gamma$ line profile, which would result in the overestimation of  $T_{\rm eff}$ for these stars by \citet{HG06}. We also find one proposed spectroscopic binary among this common sample, so the H$\gamma$ line profiles may be further altered by variable line blending effects. For the remaining B-type stars in this region it is possible that our temperature discrepancy is due to variable emission in unknown Be stars, unresolved binaries, clumping in the hot stellar wind, or differences in the atomic species included in our respective atmospheric models that affect hydrogen stark broadening \citep{Przybilla04}.

With the quantified disagreement in $T_{\rm eff}$ and log $g$ shown here, the anticipated effect of non-LTE atmospheres on the measurement results is more significant than initially assumed by \citet{HG06}. The effects of these overestimated temperatures and gravities also affect their determined stellar masses, their usage of log $g$ as an indicator of stellar evolutionary status, and their determined spin-down rates of B and Be stars. We find that cluster members are more evolved than indicated by \citet{HG06}. These B stars may spin down more slowly than the rates observed by \citet{HG06} and \citet{Huang10}. 

\begin{figure}[h]
\hspace{-.75in}
\includegraphics[angle=90,width=0.6\textwidth]{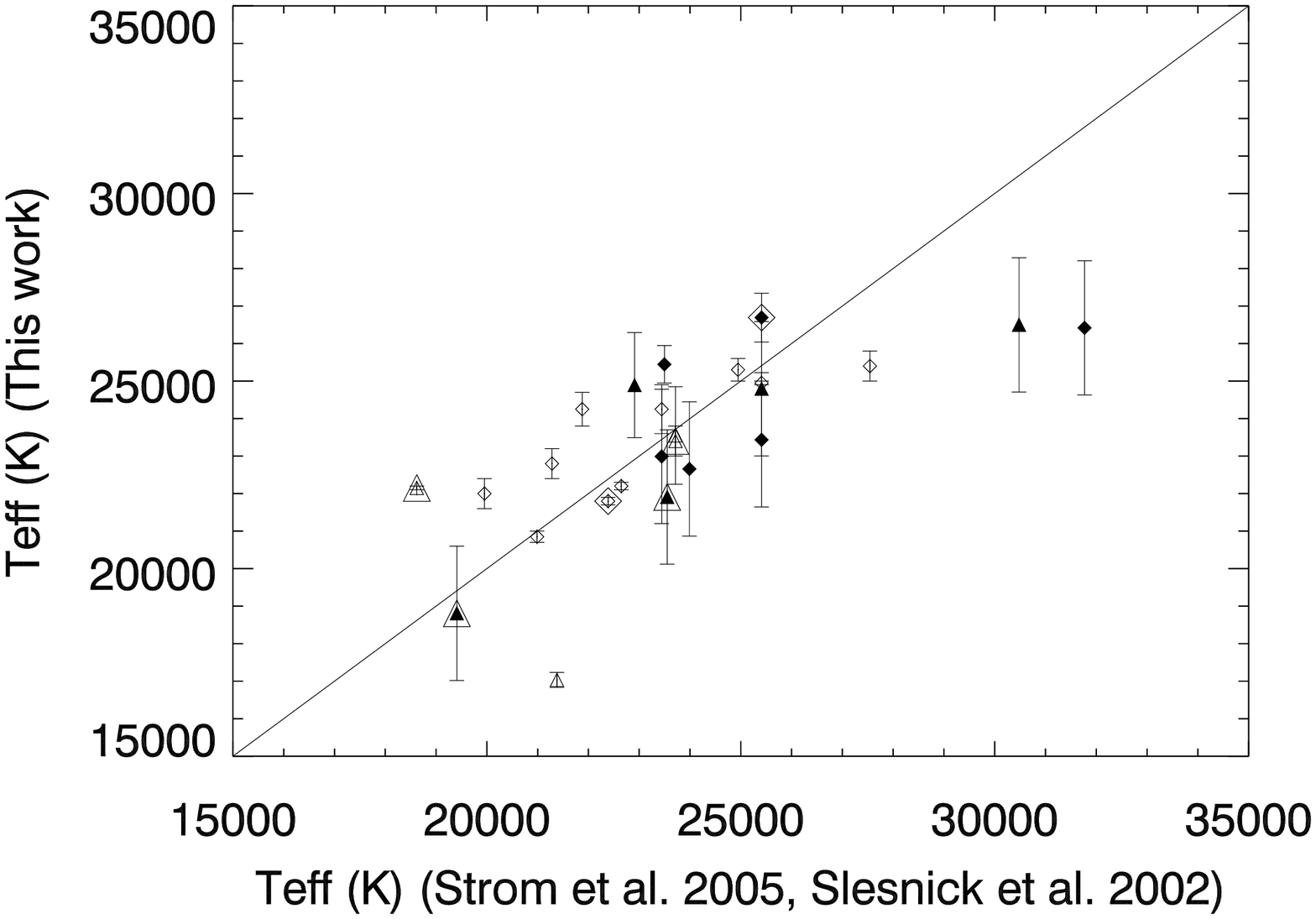}
\caption{Comparison of our $T_{\rm eff}$ measurements with those of \citet{Strom05} and \citet{Slesnick02}, in same format as Figure \ref{vsinicomp}.}
\label{StromTempComp}
\end{figure}

We also compare our results to those presented by \citet{Strom05}, who adopt the $T_{\rm eff}$ values derived from UBV photometry by \citet{Slesnick02}. In their study they find reasonable agreement between their measurements of $V$ sin $i$ and those of \citet{HG06}, though their results are systematically 5\% smaller than the results of \citet{HG06}. Comparing the results of the 26 stars common to our two samples, we find a similar agreement and systematic underestimation of Strom et al.\ 's $V$ sin $i$ values when compared to our measurements, as is expected given the excellent agreement of our results with those of \citet{HG06}. Upon further comparison of our results, we find that \citet{Strom05} and \citet{Slesnick02} have overestimated $T_{\rm eff}$ for hotter stars as well. We note that the two most discrepant stars are both Be stars, suggesting that their H$\beta$ emission contaminates the B-band brightness used to derive $T_{\rm eff}$. Thus $M_{\star}$ and $R_{\star}$ for the Be stars as shown in Figure \ref{StromTempComp} are also likely overestimated. 

Finally, we note that many of our Be stars were found to be possible spectroscopic binaries by \citet{HG06} and \citet{Strom05}. Since \citet{HG06} did not present measurements of $T_{\rm eff}$ or log $g$ for many of their spectroscopic binaries, not all of our measurements could be directly compared. Their classification as binaries may be inaccurate due to variable emission in their spectral lines. Further monitoring of their radial velocities as well as their emission will clarify their status. 

\section{Conclusions and Future Work}
We have measured $V$ sin $i$, $T_{\rm eff}$, log $g_{\rm polar}$, $M_{\star}$, and $R_{\star}$ for 104 B-type and Be star members of NGC 869 and NGC 884 using spectroscopic modeling techniques and calculations from Str\"{o}mgren photometry. Our determined values for $V$ sin $i$ are in good agreement with the earlier results of \citet{HG06}, though there is some discrepancy in our measured temperatures and surface gravities due to our use of the more recently available non-LTE BSTAR2006 stellar models of \citet{LH07} and the possible contamination of Be stars  and spectroscopic binaries. Because of the resulting over-estimation of log $g$, \citet{HG06} have underestimated the retention of initial angular momentum by the cluster members. 

We find that the cluster members are significantly more evolved than found by previous measurements. We also identify 8 transient Be stars in $h$ and $\chi$ Per. The Be stars in these clusters are also rotating more slowly than expected based upon other young open clusters. Further monitoring  of the massive stellar constituents of these clusters, and their rotation rates is well warranted.

In a forthcoming paper we will examine our sample of Be stars and their disk structures in greater detail. We will use the SEDs of B-type cluster members to compare our derived distances with previous measurements. Using our determined stellar parameters, we will be able to separate and examine the stellar and disk contributions to the total Be star + disk system flux through their observed SEDs. From these SEDs and multiple observations of H$\alpha$ we will investigate the Be disk radii, masses, and longterm variability. 

\acknowledgments
We are grateful to the anonymous referee for comments that greatly improved this manuscript. We would like to thank Yale University for providing access to the WIYN telescope at KPNO. We are grateful to the University of Wyoming for providing observing time at the WIRO 2.3m telescope, as well as to Chip Kobulnicky and Dan Kiminki for their support of our observations. We are very grateful to the support staff at NOAO, especially Di Harmer and Daryl Wilmarth for their help in obtaining our observations at KPNO. We are also very appreciative of the helpful feedback regarding this work provided by Wenjin Huang and Doug Gies. This research has made use of the WEBDA database, operated at the Institute for Astronomy of the University of Vienna. This work has been supported by the National Science Foundation under grant AST-1109247 and by NASA under DPR numbers NNX08AX79G, NNX09AP86G, and NNX11AO41G. M.V.M.\ was supported by an NSF Astronomy and Astrophysics Postdoctoral Fellowship under award AST-0401460. C.A.\ and A.N.M.B.\ are grateful for thesis student travel support provided by NOAO. A.N.M.B.\ is also supported by a Grant-In-Aid of Research (G20110315157195) from the National Academy of Sciences, administered by Sigma Xi. We are very grateful for National Science Foundation support provided for B.O.\ through REU site grant PHY-0353620 to Lehigh University. Institutional support was provided by Lehigh University.

{\it Facilities:} \facility{WIRO ()}, \facility{KPNO:CFT ()}, \facility{KPNO:2.1m ()}, \facility{WIYN ()}\\


\begin{thebibliography}{39}
\expandafter\ifx\csname natexlab\endcsname\relax\def\natexlab#1{#1}\fi

\bibitem[{Adams \& VanMaanen(1913)}]{Adams1913}
Adams, W.~S., \& VanMaanen, A. 1913, AJ, 27, 187

\bibitem[{Balona(1984)}]{Balona84}
Balona, L.~A. 1984, MNRAS, 211, 973

\bibitem[{Balona \& Shobbrook(1984)}]{BS84}
Balona, L.~A., \& Shobbrook, R.~R. 1984, MNRAS, 211, 375

\bibitem[{Bragg \& Kenyon(2002)}]{BK02}
Bragg, A., \& Kenyon, S. 2002, AJ, 124, 3289

\bibitem[{Bragg \& Kenyon(2005)}]{BK05}
---. 2005, AJ, 130, 134

\bibitem[{Cranmer(2009)}]{Cranmer09}
Cranmer, S.~R. 2009, ApJ, 701, 396

\bibitem[{Currie {et~al.}(2008)}]{Currie08}
Currie, T., {et~al.} 2008, ApJ, 672, 558

\bibitem[{Currie {et~al.}(2010)}]{Currie10}
---. 2010, ApJS, 186, 191

\bibitem[{Ekstr\"{o}m {et~al.}(2012)}]{Ekstrom12}
Ekstr\"{o}m, S., {et~al.} 2012, A\&A, 537, 146

\bibitem[{Emilio {et~al.}(2010)}]{Emilio10}
Emilio, M., {et~al.} 2010, A\&A, 522, 43

\bibitem[{Fabregat {et~al.}(1994)Fabregat, Torrej\'{o}n, \&
  Bernabeu}]{Fabregat94}
Fabregat, J., Torrej\'{o}n, J.~M., \& Bernabeu, G. 1994, Be Star Newsl, 29, 8

\bibitem[{Hertzsprung(1922)}]{Hertzsprung22}
Hertzsprung, E. 1922, BAN, 1, 151

\bibitem[{Huang \& Gies(2006{\natexlab{a}})}]{HG06}
Huang, W., \& Gies, D.~R. 2006{\natexlab{a}}, ApJ, 648, 591

\bibitem[{Huang \& Gies(2006{\natexlab{b}})}]{HG06a}
---. 2006{\natexlab{b}}, ApJ, 648, 580

\bibitem[{Huang {et~al.}(2010)Huang, Gies, \& McSwain}]{Huang10}
Huang, W., Gies, D.~R., \& McSwain, M.~V. 2010, ApJ, 722, 605

\bibitem[{Keller {et~al.}(2001)Keller, Grebel, Miller, \& Yoss}]{Keller01}
Keller, S.~C., Grebel, E.~K., Miller, G.~J., \& Yoss, K.~M. 2001, AJ, 122, 248

\bibitem[{Krzesi\'{n}ski \& Pigulski(1997)}]{Krzesinski97}
Krzesi\'{n}ski, J., \& Pigulski, A. 1997, A\&A, 325, 987

\bibitem[{Krzesi\'{n}ski {et~al.}(1999)Krzesi\'{n}ski, Pigulski, \&
  Ko{\l}aczkowski}]{Krzesinski99}
Krzesi\'{n}ski, J., Pigulski, A., \& Ko{\l}aczkowski, Z. 1999, A\&A, 345, 505

\bibitem[{{Kurucz}(1994)}]{Kurucz94}
{Kurucz}, R. 1994, Solar abundance model atmospheres for 0,1,2,4,8 km/s.~Kurucz
  CD-ROM No.~19.~ Cambridge, Mass.: Smithsonian Astrophysical Observatory,
  1994., 19

\bibitem[{Lanz \& Hubeny(2007)}]{LH07}
Lanz, T., \& Hubeny, I. 2007, ApJS, 169, 83

\bibitem[{McSwain {et~al.}(2009)McSwain, Huang, \& Gies}]{McSwain09}
McSwain, M.~V., Huang, W., \& Gies, D.~R. 2009, ApJ, 700, 1216

\bibitem[{McSwain {et~al.}(2008)McSwain, Huang, Gies, Grundstrom, \&
  Townsend}]{McSwain08}
McSwain, M.~V., Huang, W., Gies, D.~R., Grundstrom, E.~D., \& Townsend, R.
  H.~D. 2008, ApJ, 672, 590

\bibitem[{Messow(1913)}]{Messow1913}
Messow, B. 1913, AAHam, 2b, 1M

\bibitem[{Meynet \& Maeder(2000)}]{Meynet00}
Meynet, G., \& Maeder, A. 2000, A\&A, 361, 101

\bibitem[{Napiwotzki {et~al.}(1993)Napiwotzki, Schoenberner, \&
  Wenske}]{Napiwotzki93}
Napiwotzki, R., Schoenberner, D., \& Wenske, V. 1993, A\&A, 268, 653

\bibitem[{Porter \& Rivinius(2003)}]{Porter03}
Porter, J.~M., \& Rivinius, T. 2003, PASP, 115, 1153

\bibitem[{Przybilla \& Butler(2004)}]{Przybilla04}
Przybilla, N., \& Butler, K. 2004, ApJ, 609, 1181

\bibitem[{Przybilla {et~al.}(2011)Przybilla, Nieva, \& Butler}]{Przybilla11}
Przybilla, N., Nieva, M., \& Butler, K. 2011, JPhCS, 328, 012015

\bibitem[{Rivinius {et~al.}(2003)Rivinius, Baade, \& \u{S}tefl}]{Rivinius03}
Rivinius, T., Baade, D., \& \u{S}tefl, S. 2003, A\&A, 411, 229

\bibitem[{Rivinius {et~al.}(2001)}]{Rivinius01}
Rivinius, T., {et~al.} 2001, A\&A, 369, 1058

\bibitem[{Saesen {et~al.}(2010)}]{Saesen10}
Saesen, S., {et~al.} 2010, A\&A, 515, A16

\bibitem[{Schaller {et~al.}(1992)Schaller, Schraerer, Meynet, \&
  Maeder}]{Schaller92}
Schaller, G., Schraerer, D., Meynet, G., \& Maeder, A. 1992, A\&AS, 96, 269

\bibitem[{Schild(1966)}]{Schild66}
Schild, R.~E. 1966, ApJ, 146, 142

\bibitem[{Slesnick {et~al.}(2002)Slesnick, Hillenbrand, \& Massey}]{Slesnick02}
Slesnick, C., Hillenbrand, L.~A., \& Massey, P. 2002, ApJ, 576, 880

\bibitem[{Slettebak(1968)}]{Slettebak68}
Slettebak, A. 1968, ApJ, 154, 933

\bibitem[{Slettebak(1985)}]{Slettebak85}
---. 1985, ApJ, 59, 769

\bibitem[{Southworth {et~al.}(2004)Southworth, Zucker, Maxted, \&
  Smalley}]{Southworth04}
Southworth, J., Zucker, S., Maxted, P. F.~L., \& Smalley, B. 2004, MNRAS, 355,
  986

\bibitem[{Strom {et~al.}(2005)Strom, Wolff, \& Dror}]{Strom05}
Strom, S.~E., Wolff, S.~C., \& Dror, D. H.~A. 2005, AJ, 129, 809

\bibitem[{Trumpler(1926)}]{Trumpler26}
Trumpler, R.~J. 1926, PASP, 38, 350

\end{thebibliography}

\clearpage

\clearpage

\begin{landscape}
\begin{table}[h]
\begin{center}
\caption{Journal of Spectroscopy}
\scriptsize
\begin{tabular}{lccclcccc}\tableline
&{\bf Range}&{\bf Resolving Power}&{\bf Number of}& {\bf Telescope/}&{\bf Grating/}&&{\bf Camera/}\\
{\bf UT Dates}&{\bf (\AA)}&{\bf ($\lambda / \Delta\lambda$)}&{\bf Targets}&{\bf  Spectrograph}&{\bf Order}&{\bf Filter}&{\bf Detector}\\ \hline
2005 Nov 14--15& 4250--4900& 4700 & 92 & WIYN 3.5m / Hydra & 1200@28.7/2& BG39&Red bench, blue fibers / 5TA1\\
2010 Aug 23--27 &4000--5200& 1600 & 2&  WIRO 2.3m / Long Slit & LS-1/2& BG40&--- / SBIG-ST-2000 \\ 
2010 Aug 28--31 &5350--6810&4500&6 & WIRO 2.3m / Long Slit & LS-2/1&GG455&--- / SBIG-ST-2000\\ 
2011 Nov 10--13 &4000--5000&2300 &1 & KPNO 2.1m/ GoldCam&47/2&CuSO4, WG354&Gold / F3KC\\
2012 Jan 3--8 &4090--4550&7800&2& KPNO CF/Coud\'e &B/3&4-96&Cam5 / T2KB\\
2012 Jan 3--8 &6320--7025&11700&1& KPNO CF/Coud\'e &B/2&OG-55&Cam5 / T2KB\\\tableline
 \end{tabular}
 \label{spectroscopy}
 \end{center}
 \vspace{1mm}
 \scriptsize
\end{table}
\end{landscape}


\begin{deluxetable}{ccccccccccccl}
\tabletypesize{\scriptsize}
\tablecolumns{13}
\rotate
\tablecaption{Physical Parameters of B-type Cluster Members \label{starparams}}
\tablewidth{0pt}
\tablehead{
\colhead{\textbf{WEBDA}}&
\colhead{\textbf{$V$ sin $i$}}&
\colhead{\textbf{$\Delta V$ sin $i$}}&
\colhead{\textbf{$T_{\rm eff}$}}&
\colhead{\textbf{$\Delta T_{\rm eff}$}}&
\colhead{\textbf{log $g$}}&
\colhead{\textbf{$\Delta$ log $g$}}&
\colhead{\textbf{log $g_{\rm polar}$}}&
\colhead{\textbf{$M_{\star}$}}&
\colhead{\textbf{$\Delta M_{\star}$}}&
\colhead{\textbf{$R_{\star}$}}&
\colhead{\textbf{$\Delta R_{\star}$}}\\
\colhead{\textbf{ID}}&
\colhead{\textbf{(km~s$^{-1}$)}}&
\colhead{\textbf{(km~s$^{-1}$)}}&
\colhead{\textbf{(K)}}&
\colhead{\textbf{(K)}}&
\colhead{\textbf{(dex)}}&
\colhead{\textbf{(dex)}}&
\colhead{\textbf{(dex)}}&
\colhead{\textbf{($M_{\odot}$)}}&
\colhead{\textbf{($M_{\odot}$)}}&
\colhead{\textbf{($R_{\odot}$)}}&
\colhead{\textbf{($R_{\odot}$)}}&\\
\colhead{(1)}&
\colhead{(2)}&
\colhead{(3)}&
\colhead{(4)}&
\colhead{(5)}&
\colhead{(6)}&
\colhead{(7)}&
\colhead{(8)}&
\colhead{(9)}&
\colhead{(10)}&
\colhead{(11)}&
\colhead{(12)}
}

\startdata
{\bf NGC 869}&&&&&&&&&&&\\
  63\tablenotemark{a}  &  151  &   8  &  21800  &   250  & 3.81  & 0.02  & 3.92  & 8.7  & 0.1  & 5.4  & 1.7  \\
  87  &   89  &  16  &  10060  &    60  & 3.83  & 0.03  & 3.92  & 2.8  & 0.1  & 3.0  & 0.3  \\
  90  &   64  &   6  &  22350  &    50  & 3.65  & 0.02  & 3.68  &10.7  & 0.1  & 7.8  & 1.4  \\
 133\tablenotemark{a}  &  341  &  19  &  17500 &   200  & 3.92  & 0.02  & 4.22  & 5.4  & 0.1  & 3.0  & 1.2  \\
 138  &   97  &  12  &  19400  &   200  & 3.93  & 0.02  & 3.99  & 6.9  & 0.1  & 4.4  & 1.3  \\
 197  &  262  &  25  &  10340  &    51  & 3.89  & 0.03  & 4.13  & 2.6  & 0.1  & 2.3  & 0.1  \\
 250  &  109  &  25  &  17750  &    50  & 4.00  & 0.02  & 4.07  & 5.9  & 0.1  & 3.7  & 1.1  \\
 260  &  198  &  13  &  25200  &   200  & 3.33  & 0.02  & 3.54  &14.9  & 0.3  &10.8  & 2.0  \\
 289  &   66  &  10  &  20100  &   100  & 3.95  & 0.02  & 3.98  & 7.4  & 0.1  & 4.6  & 1.4  \\
 323\tablenotemark{a}  &   50  &   7  &  20700  &   300  & 3.78  & 0.03  & 3.80  & 8.6  & 0.2  & 6.1  & 1.2  \\
 339  &   76  &   7  &  27000  &  1000  & 3.55  & 0.07  & 3.58  &17.2  & 1.5  &11.1  & 3.0  \\
 350  &  136  &   4  &  18450  &    50  & 3.90  & 0.02  & 4.00  & 6.4  & 0.1  & 4.2  & 1.1  \\
 380  &  112  &   6  &  19150  &    50  & 4.00  & 0.02  & 4.07  & 6.6  & 0.1  & 3.9  & 1.3  \\
 530\tablenotemark{a}  &   59  &   9  &  24100  &   350  & 3.95  & 0.03  & 3.97  &10.5  & 0.3  & 5.5  & 2.4  \\
 551\tablenotemark{a}  &  220  &  11  &  20200  &   350  & 3.63  & 0.03  & 3.83  & 8.1  & 0.2  & 5.7  & 1.2  \\
 590  &  119  &  10  &  24050  &   50  & 3.75  & 0.02  & 3.82  &11.2  & 0.2  & 6.8  & 2.2  \\
 662  &   67  &  11  &  24300  &   650  & 3.13  & 0.05  & 3.17  &20.0  & 1.7  &19.2  & 1.4  \\
 678  &  100  &  15  &  22950  &   50  & 3.60  & 0.02  & 3.67  &11.3  & 0.1  & 8.1  & 1.6  \\
 692  &  176  &  16  &  25400  &   400  & 3.38  & 0.03  & 3.55  &15.0  & 0.6  &10.7  & 2.1  \\
 731\tablenotemark{a}  &  165  &  11  &  21600  &   250  & 3.94  & 0.02  & 4.05  & 8.2  & 0.1  & 4.5  & 1.8  \\
 748  &  251  &  35  &  10900  &    78  & 3.61  & 0.03  & 3.89  & 3.1  & 0.1  & 3.3  & 0.2  \\
 768  &   45  &   6  &  17600  &   200  & 3.48  & 0.02  & 3.51  & 7.7  & 0.1  & 8.1  & 0.4  \\
 769  &  127  &   8  &  22650  &    50  & 3.75  & 0.02  & 3.84  & 9.9  & 0.1  & 6.3  & 1.8  \\
 789  &  199  &  16  &  17250  &    50  & 4.00  & 0.02  & 4.15  & 5.4  & 0.1  & 3.2  & 1.1  \\
 803  &  265  &  19  &  23100  &   150  & 3.49  & 0.03  & 3.75  &10.9  & 0.1  & 7.3  & 1.8  \\
 843\tablenotemark{a}  &  113  &   7  &  21800  &   100  & 3.40  & 0.02  & 3.50  &11.4  & 0.1  & 9.9  & 0.8  \\
 859  &  221  &   9  &  22000  &   400  & 3.73  & 0.02  & 3.91  & 8.9  & 0.2  & 5.5  & 1.7  \\
 864  &  150  &  45  &  22800  &   400  & 3.73  & 0.02  & 3.84  &10.1  & 0.4  & 6.3  & 1.9  \\
 922  &  270  &  14  &  25300  &   300  & 3.34  & 0.02  & 3.62  &14.2  & 0.3  &9.6  & 2.3  \\
1001  &   61  &   4  &  22900  &   100  & 3.60  & 0.02  & 3.63  &11.5  & 0.1  & 8.6  & 1.4  \\
1067\tablenotemark{a}  &   63  &   8  &  21800  &   300  & 3.72  & 0.02  & 3.75  & 9.6  & 0.3  & 6.8  & 1.4  \\
1078  &  208  &  24  &  24950  &    50  & 3.50  & 0.02  & 3.69  &13.1  & 0.1  & 8.6  & 2.2  \\
1085  &   88  &   5  &  22200  &   100  & 3.50  & 0.02  & 3.57  &11.3  & 0.1  & 9.2  & 1.1  \\
1132  &  103  &   7  &  24300  &   450  & 3.18  & 0.02  & 3.27  &17.9  & 0.8  &16.2  & 0.9  \\
1141  &  233  &  14  &  20700  &   150  & 3.20  & 0.02  & 3.49  &10.4  & 0.1  & 9.6  & 0.4  \\
1174  &   65  &   9  &  19300  &   200  & 3.93  & 0.02  & 3.96  & 6.9  & 0.1  & 4.5  & 1.2  \\
1181  &   71  &   6  &  17300  &   200  & 3.97  & 0.03  & 4.01  & 5.9  & 0.1  & 4.0  & 1.0  \\
1364  &  235  &  12  &  20900  &   150  & 3.40  & 0.02  & 3.66  & 9.4  & 0.1  & 7.5  & 0.9  \\
1387  &   79  &   4  &  20500  &   100  & 3.85  & 0.02  & 3.90  & 8.1  & 0.1  & 5.3  & 1.4  \\
1391  &  161  &   9  &  22300  &   100  & 3.55  & 0.02  & 3.69  &10.6  & 0.1  & 7.7  & 1.4  \\
1482  &  101  &  59  &  10740  &    72  & 3.47  & 0.03  & 3.61  & 3.5  & 0.0  & 4.9  & 0.9  \\
1516\tablenotemark{a}  &  152  &  17  &  25400  &   650  & 4.24  & 0.05  & 4.29  & 9.9  & 0.5  & 3.7  & 0.2  \\
1548  &  195  &  14  &  13300  &   171  & 4.12  & 0.04  & 4.23  & 3.5  & 0.1  & 2.4  & 0.6  \\

\bf{NGC 884}&&&&&&&&&&&&\\
1770  &  232  &   7  &  17650  &    50  & 3.75  & 0.02  & 3.97  & 6.2  & 0.1  & 4.3  & 1.0  \\
1793  &  136  &  30  &      --  &     --  & --  & --  & --  & --  & --  & --  & --  \\
1873\tablenotemark{a}  &  166  &   8  &  21600  &   300  & 3.83  & 0.02  & 3.95  & 8.5  & 0.2  & 5.1  & 1.7  \\
1899  &  137  &   7  &  23500  &   700  & 3.16 & 0.05  & 3.31  &15.9  & 1.3  &14.6  & 0.6  \\
2014\tablenotemark{a}  &  293  &   8  &  19000  &   250  & 3.64  & 0.03  & 3.93  & 6.9  & 0.1  & 4.7  & 1.1  \\
2048  &  100  &   6  &  19400  &   250  & 3.92  & 0.02  & 3.98  & 6.9  & 0.1  & 4.4  & 1.2  \\
2049  &  413  &  28  &  18050  &    50  & 2.75  & 0.02  & 3.61  & 7.5  & 0.1  & 7.1  & 0.2  \\
2057  &  151  &  10  &  19450  &   50  & 3.80  & 0.02  & 3.92  & 7.2  & 0.1  & 4.9  & 1.1  \\
2085  &  284  &   9  &  19000  &   250  & 3.54  & 0.02  & 3.85  & 7.1  & 0.2  & 5.3  & 0.9  \\
2086  &  146  &  34  &      --  &     --  & --  & --  & --  & --  & --  & --  & --  \\
2094  &   80  &   3  &  20350  &   50  & 4.00  & 0.02  & 4.04  & 7.4  & 0.1  & 4.3  & 1.6  \\
2112  &   89  &   8  &  22300  &   150  & 3.80  & 0.02  & 3.85  & 9.4  & 0.2  & 6.0  & 1.7  \\
2119  &   50  &   5  &  22400  &   100  & 3.95  & 0.02  & 3.97  & 8.9  & 0.1  & 5.1  & 1.9  \\
2185  &  202  &  12  &  17000  &   200  & 3.93  & 0.02  & 4.10  & 5.5  & 0.1  & 3.5  & 1.0  \\
2190  &  335  &  28  &  20400  &   450  & 3.18  & 0.02  & 3.67  & 8.9  & 0.3  & 7.2  & 0.8  \\
2191\tablenotemark{c}  &  263  &  16  &  18050  &    50  & 3.80  & 0.02  & 4.04  & 6.2  & 0.1  & 3.9  & 1.1  \\
2218  &  248  &  35  &  10110  &    57  & 3.84  & 0.03  & 4.07  & 2.6  & 0.1  & 2.5  & 0.1  \\
2227  &   62  &   7  &  23600  &  1300  & 3.00  & 0.10  & 3.07  &21.0  & 4.0  &22.2  & 4.8  \\
2232\tablenotemark{c}  &  115  &   5  &  22150  &   50  & 3.90  & 0.02  & 3.97  & 8.8  & 0.1  & 5.1  & 1.8  \\
2255  &  333  &  15  &  20200  &   150  & 3.38  & 0.02  & 3.81  & 8.2  & 0.1  & 5.9  & 1.1  \\
2311\tablenotemark{b}  &   51  &   5  &  23400  &   400  & 3.48  & 0.02  & 3.50  &13.3  & 0.5  &10.8  & 1.3  \\
2336\tablenotemark{a}  &  173  &   9  &  19200  &   200  & 3.92  & 0.02  & 4.05  & 6.7  & 0.1  & 4.0  & 1.3  \\
2347  &   84  &  11  &  21800  &   100  & 4.00  & 0.02  & 4.04  & 8.3  & 0.1  & 4.6  & 1.9  \\
2361  &   68  &   8  &  25500  &   500  & 3.37  & 0.05  & 3.40  &17.7  & 0.8  &13.9  & 1.9  \\
2392  &   87  &  17  &  21200  &   150  & 3.75  & 0.02  & 3.81  & 8.8  & 0.1  & 6.1  & 1.3  \\
2520  &   45  &  10  &  23200  &   350  & 3.67  & 0.02  & 3.68  &11.4  & 0.3  & 8.0  & 1.7  \\
2540  &  271  &  23  &  19800  &   250  & 3.53  & 0.02  & 3.82  & 7.9  & 0.2  & 5.7  & 1.1  \\
2555  &  129  &   6  &  18250  &    50  & 4.05  & 0.02  & 4.13  & 6.0  & 0.1  & 3.5  & 1.3  \\
2605  &   54  &  27  &  24600  &   450  & 3.58  & 0.02  & 3.60  &13.6  & 0.5  & 9.7  & 1.9  \\
2622  &  373  &  35  &  17200  &   200  & 3.38  & 0.03  & 3.91  & 6.1  & 0.1  & 4.6  & 0.8  \\
2716\tablenotemark{a}  &  198  &   7  &  19650  &   50  & 3.75  & 0.02  & 3.92  & 7.3  & 0.1  & 4.9  & 1.2  \\
2729  &  208  &  17  &  10410  &    74  & 3.83  & 0.03  & 4.01  & 2.8  & 0.1  & 2.7  & 0.1  \\
2907  &  233  &  52  &  13700  &   150  & 3.86  & 0.03  & 4.07  & 3.9  & 0.1  & 3.0  & 0.4  \\
\enddata
\tablenotetext{a}{Proposed spectroscopic binary from \citet{HG06a}}
\tablenotetext{b}{Eclipsing binary from \citet{Southworth04}}
\tablenotetext{c}{Proposed eclipsing binary from \citet{Saesen10}}
\end{deluxetable}



\begin{deluxetable}{ccccccccccccl}
\tabletypesize{\scriptsize}
\tablecolumns{13}
\rotate
\tablecaption{Physical Parameters of Be Star Cluster Members \label{Be_params}}
\tablewidth{0pt}
\tablehead{
\colhead{\textbf{WEBDA}}&
\colhead{\textbf{$V$ sin $i$}}&
\colhead{\textbf{$\Delta V$ sin $i$}}&
\colhead{\textbf{$T_{\rm eff}$}}&
\colhead{\textbf{$\Delta T_{\rm eff}$}}&
\colhead{\textbf{log $g$}}&
\colhead{\textbf{$\Delta$ log $g$}}&
\colhead{\textbf{log $g_{\rm polar}$}}&
\colhead{\textbf{$M_{\star}$}}&
\colhead{\textbf{$\Delta M_{\star}$}}&
\colhead{\textbf{$R_{\star}$}}&
\colhead{\textbf{$\Delta R_{\star}$}}&
\colhead{\textbf{Notes}} \\
\colhead{\textbf{ID}}&
\colhead{\textbf{(km~s$^{-1}$)}}&
\colhead{\textbf{(km~s$^{-1}$)}}&
\colhead{\textbf{(K)}}&
\colhead{\textbf{(K)}}&
\colhead{\textbf{(dex)}}&
\colhead{\textbf{(dex)}}&
\colhead{\textbf{(dex)}}&
\colhead{\textbf{($M_{\odot}$)}}&
\colhead{\textbf{($M_{\odot}$)}}&
\colhead{\textbf{($R_{\odot}$)}}&
\colhead{\textbf{($R_{\odot}$)}}&\\
\colhead{(1)}&
\colhead{(2)}&
\colhead{(3)}&
\colhead{(4)}&
\colhead{(5)}&
\colhead{(6)}&
\colhead{(7)}&
\colhead{(8)}&
\colhead{(9)}&
\colhead{(10)}&
\colhead{(11)}&
\colhead{(12)}&
\colhead{(13)}
}
\startdata
{\bf NGC 869}&&&&&&&&&&&&\\
  49  &  172  &  21  &  25400  &  1790  & 3.54  & 0.23  & 3.54  &15.4  & 2.5  & 11.0  & 2.0  &No Emission Observed(1);\\
  		&&&&&&&&&&&&							Emission Observed(2,7,8)\\
 146  &  195  &   6  &  25600  &   600  & 3.66  & 0.05  & 3.80  &12.7  & 0.7  & 7.4  & 2.6  &No Emission Observed(2,8); \\
 &&&&&&&&&&&&									Emission Observed(4)\\
 309  &    --  &   --  &  27700  &  1790  & 3.74  & 0.23  & 3.74  &15.8  & 1.6  & 8.8  & 2.1  &Emission Observed(1,2,4,5,8)\\
 517  &  178  &  13  &      --  &     --  & --  & --  & --  & --  & --  & --  & --  &Emission Observed(4,6,7,8)\\
 566\tablenotemark{a}  &  306  &  25  &  22100  &   100  & 3.25  & 0.02  & 3.63  &10.8  & 0.1  & 8.4  & 1.2  &No Emission Observed(1,4,8);\\
 &&&&&&&&&&&&								Emission Observed(2,7,8)\\
 717\tablenotemark{b}  &  126  &  22  &  26700  &   650  & 3.91  & 0.05  & 3.96  &12.5  & 0.8  & 6.1  & 3.2  &No Emission Observed(6,8);\\
 &&&&&&&&&&&&								 Emission Observed(3)\\
 846  &  205  &  19  &      --  &     --  & --  & --  & --  & --  & --  & --  & --  &Emission Observed(4,6,7,8)\\
 847  &   87  &  11  &  23000  &  1790  & 3.41  & 0.24  & 3.41  &13.7  & 0.2  &12.1  & 2.8  &Emission Observed(2,4,6,7,8)\\
 992  &    --  &   --  &  21700  &  1790  & 3.60  & 0.24  & 3.60  &10.7  & 0.1  & 8.6  & 2.0  &Emission Observed(6,8)\\
1057  &    --  &   --  &  21300 &  1790  & 2.69  & 0.25  & 2.69  &29.5  &0.2  &40.8  & 9.6  &Emission Observed(6,8)\\
1161  &    --  &   --  &  23400  &  1790  & 3.77  & 0.24  & 3.77  &11.1  & 0.2  & 7.2  & 1.7  &Emission Observed(2,4,5,6,7,8)\\
1261  &  285  &  78  &  26400  &  1790  & 3.72  & 0.23  & 3.72  &14.3  & 0.6  & 8.6  & 2.0  &Emission Observed(2,4,5,6,7,8)\\
1268  &  151  &  12  &  25400  &   500  & 3.46  & 0.05  & 3.59  &14.6  & 0.6  &10.1  & 2.2  &No Emission Observed(6,8);\\
 &&&&&&&&&&&&								Emission Observed(4)\\
1278  &  197  &  12  &  25100  &  1790  & 4.24  & 0.24  & 4.24  & 9.9  & 0.5  & 3.9  & 0.9  &Emission Observed(4,6,7,8)\\
1282  &    --  &   --  &  22700  &  1790  & 3.95  & 0.24  & 3.95  & 9.2  & 0.2  & 5.3  & 1.2  &Emission Observed(2,4,5,6,7,8)\\
 \\
\bf{NGC 884}&&&&&&&&&&&&\\
1702\tablenotemark{a}  &     -- & -- &       24400 &    1790  & 3.67  & 0.24  & 3.67  & 12.6  & 0.3  & 8.6  & 2.0  & Emission Observed(1,2,4,6,7,8)\\
1772  &  379  &  28  &      --  &     --  & --  & --  & --  & --  & --  & --  & --  &No Emission Observed(4);\\
 &&&&&&&&&&&&								Emission Observed(6,8)\\
1926\tablenotemark{a}  &  106  &  17  &  27400  &  1790  & 3.89  & 0.23  & 3.89  &13.9  & 1.2  & 7.0  & 1.6  &Emission Observed(2,4,6,7,8)\\
1977  &     -- & -- &      20400  &     1790 & 4.27  & 0.25  & 4.27  & 6.8  & 0.1  & 3.2  & 0.7  & Emission Observed(2,4,6,7,8)\\
2088\tablenotemark{a}  &    --  &   --  &  21900 &  1790  & 3.72  & 0.24  & 3.72  & 10.0  & 0.1  & 7.2  & 1.7  &Emission Observed(2,4,5,6,7,8)\\
2091\tablenotemark{a}  &  236  &  40  &  18800  &  1790  & 3.97  & 0.26  & 3.97  & 6.7  & 0.1  & 4.4  & 1.0  &Emission Observed(4,6,7,8)\\
2138  &  153  &  37  &  24100  &  1790  & 3.53  & 0.24  & 3.53  &13.8  & 0.2  & 10.5  & 2.5  &Emission Observed(1,2,4,5,7,8)\\
2165  &    --  &   --  &  24800  &  1790  & 3.79 & 0.24  & 3.79  &11.9  & 0.3  & 7.3  & 1.7  & Emission Observed(2,4,5,6,7,8)\\
2262  &  265  &  19  &  24900  &  1400  & 3.51  & 0.08  & 3.75  &12.3  & 1.4  & 7.7  & 2.2  &No Emission Observed(4,8);\\
&&&&&&&&&&&&													Emission Observed(2,6)\\
2284  &    --  &   --  &  26500  &  1790  & 3.77  & 0.23  & 3.76  &14.0  & 0.7  & 8.1  & 1.9  &Emission Observed(2,4,5,6,7,8)\\
2468  &  134  &  46  &  10070  &    48  & 3.88  & 0.03  & 4.00  & 2.7  & 0.1  & 2.7  & 0.1  &No Emission Observed(8);\\
 &&&&&&&&&&&&								 Emission Observed(7)\\
2563  &  308  &  74  &  26100  &  1790  & 4.02  & 0.23  & 4.02  &11.6  & 1.5  & 5.5  & 2.7  &Emission Observed(2,4,6,7,8)\\
2771  &    --  &   --  &  21800  &  1790  & 3.99  & 0.24  & 3.99  & 8.5  & 0.1  & 4.9  & 1.1  &Emission Observed(2,4,7,8)\\
\enddata
\\
References: (1) \citet{Schild66}, (2) \citet{Slettebak85}, (3) \citet{Fabregat94}, (4) \citet{Keller01}, (5) \citet{Slesnick02}, (6) \citet{BK02}, (7) Be candidate with observed IR excesses in \citet{Currie08}, (8) This work.\\

\tablenotetext{a}{Proposed spectroscopic binary from \citet{HG06a}}
\tablenotetext{b}{Candidate spectroscopic binary from \citet{Strom05}}

\end{deluxetable}

\clearpage

\end{document}